\documentclass[12pt,a4paper,final]{iopart}

\usepackage{iopams}  
\usepackage{graphicx}
\usepackage[breaklinks=true,colorlinks=true,linkcolor=blue,urlcolor=blue,citecolor=blue]{hyperref}

\begin{document}

\title[Electronic properties of $ LaAlO_3 $/$ SrTiO_3 $ n-type interfaces: 
       A GGA+$ U $ study]
      {Electronic properties of $ {\bf LaAlO_3} $/$ {\bf SrTiO_3} $ n-type 
       interfaces: A GGA+$ U $ study}

\author{I.~I.~Piyanzina$^{1,2}$}
\address{$^1$Institute of Physics, Kazan Federal University, 
             Kremlyovskaya St.~18, 420008 Kazan, Russia}
\address{$^2$EP VI and Center for Electronic Correlations and Magnetism, 
             Universit\"at Augsburg, Universit\"atsstra{\ss}e~1, 
             86135 Augsburg, Germany}
\ead{irina.piyanzina@physik.uni-augsburg.de}

\author{T.~Kopp}
\address{EP VI and Center for Electronic Correlations and Magnetism, 
         Universit\"at Augsburg, Universit\"atsstra{\ss}e~1, 
         86135 Augsburg, Germany}

\author{Yu.~V.~Lysogorskiy}
\address{Institute of Physics, Kazan Federal University, 
         Kremlyovskaya St.~18, 420008 Kazan, Russia}
 
\author{D.~A.~Tayurskii}
\address{Institute of Physics, Kazan Federal University, 
         Kremlyovskaya St.~18, 420008 Kazan, Russia}
 
\author{V.~Eyert}
\address{Materials Design SARL, 42 Avenue Verdier, 92120 Montrouge, France}

\begin{abstract}
The r\^{o}le of electronic correlation effects for a realistic description 
of the electronic properties of $ {\rm LaAlO_3} $/$ {\rm SrTiO_3} $ 
heterostructures as covered by the on-site Coulomb repulsion within 
the GGA+$ U $ approach is investigated. Performing a systematic variation 
of the values of the Coulomb parameters applied to the Ti $ 3d $ and La $ 4f $ 
orbitals we put previous suggestions to include a large value for the 
La $ 4f $ states into perspective. Furthermore, our calculations provide 
deeper insight into the band gap landscape in the space spanned by these 
Coulomb parameters and the resulting complex interference effects. In 
addition, we identify important correlations between the local Coulomb 
interaction within the La $ 4f $ shell, the band gap, and the atomic 
displacements at the interface. In particular, these on-site Coulomb 
interactions influence buckling within the LaO interface layer, which via 
its strong coupling to the electrostatic potential in the LAO overlayer 
causes considerable shifts of the electronic states at the surface 
and eventually controls the band gap. 
\end{abstract}

\submitto{J.\ Phys.: Condens.\ Matter}
\maketitle

\section{Introduction}

The metallic heterointerface between two insulating oxides~ \cite{ohtomo2004}, 
namely, polar $ {\rm LaAlO_3} $ (LAO) and non-polar $ {\rm SrTiO_3} $ (STO) 
is continuously attracting high interest (see, for example, the recent reviews
\cite{bristowe2014,gariglio2015}). The 
metallic conductivity occurs when the LAO overlayers reach a critical 
thickness and the impending polar catastrophe is mitigated by charge 
transfer from the LAO surface to the STO side of the interface~\cite{nakagawa2006,thiel2006}. 
However, the number of LAO layers required to generate a metallic 
state is still a matter of debate \cite{chen2010}. While experimentally 
a metallic interface state was found when the LAO film reaches a critical 
thickness of four unit cells \cite{thiel2006}, {\it ab initio} calculations 
as based on density functional theory (DFT) led to differing results. 

This is to some extent due to the use of different functionals within 
DFT. In particular, as well known, local and semilocal exchange-correlation 
functionals as provided by the local density approximation (LDA) and the 
generalized gradient approximation (GGA) underestimate the band 
gap of semiconductors and insulators and therefore cast doubt on the 
transition point. This shortcoming may be overcome by the GGA+$ U $ approach, 
which, allows to take into account local correlations within a selected 
subset of orbitals \cite{anisimov1991,anisimov1993,czyzyk1994}. However, 
results depend on the value of the on-site Coulomb interaction $ U $ (and 
the Hund's rule coupling parameter $ J $) as well as on the set of orbitals 
these corrections are applied to.  
Probably the most accurate approach currently available is provided by 
hybrid functionals. Indeed, while from GGA calculations band gaps of 
1.6~eV for STO and 3.5~eV for LAO were obtained, which are considerably 
smaller than the experimental values of 3.2~eV and 5.6~eV, respectively, 
hybrid functional calculations led to 3.1~eV and 5.0~eV \cite{eyert2010}. 
Very similar values were reported by Mitra {\em et al.} \cite{mitra2012}.
In contrast, Nazir and Yang opted for the GGA+$ U $ approach with 
$ U_{\rm La} = 7.5 $~eV and $ U_{\rm Ti} = 5.8 $~eV \cite{nazir2014}. 
Applying these values to the parent compounds they found band gaps of 
2.5~eV for STO and 3.1~eV for LAO, which are close to the GGA results.  

Using the same values of $ U_{\rm La} $ and $ U_{\rm Ti} $ in their 
calculations for a heterostructure, Nazir and Yang obtained a band gap 
of 0.15~eV for a slab with four LAO layers and metallic behaviour 
beyond. This finding is in qualitative agreement with the results of 
Cossu {\em et al.}, who used a symmetric slab with $ 5 \frac{1}{2} $ 
central STO layers sandwiched by a varying number of LAO layers with 
the slabs separated by a 20 \AA\ thick vacuum region \cite{cossu2013}. 
Applying a hybrid functional they obtained a band gap of 0.6~eV for 
four LAO layers, while the structure with five layers was found to 
be metallic. However, hybrid functional calculations are computationally 
very demanding, which is a limiting factor especially for the study of 
transition-metal oxide heterostructures with large unit cells. 
In view of the general agreement that in these systems the electronic 
correlations beyond the semilocal approximation are well captured by 
on-site Coulomb interactions, the GGA+$ U $ approach offers a viable 
alternative. In adopting this approach, Breitschaft {\em et al.}\ used  
values $ U = 2 $~eV and $ J = 0.8 $~eV at the Ti sites. In addition, 
in order to avoid a spurious mixing of the La $ 4f $ states with the 
Ti $ 3d $ bands, a large $ U $ of 8~eV was imposed on the La $ 4f $ 
orbitals \cite{breitschaft2010,pavlenko2011}. The necessity of introducing 
on-site correlations in the form of an additional Hubbard $ U $ also for 
the La $ 4f $ states to reduce their impact on the lower conduction bands 
was previously pointed out by Okamoto {\em et al.}, who used yet increased 
values of $ U_{\rm La} = 11 $~eV and $ U_{\rm Ti} = 5 $~eV as well as 
$ J_{\rm La} = 0.68 $~eV and $ J_{\rm Ti} = 0.64 $~eV \cite{okamoto2006}. 
The same values were also adopted by Zhong and Kelly \cite{zhong2008}. 
Pentcheva and Pickett used still another parameter set of 
$ U_{\rm La} = 7.5 $~eV and $ U_{\rm Ti} = 8 $~eV as well as 
$ J_{\rm Ti} = 1 $~eV \cite{pentcheva2008}. 
Nevertheless, an exhausting investigation of the dependence of the 
electronic properties on the values of $ U $ is still missing. To this 
end, the aim of the present contribution is to provide a systematic study 
of the impact of the strength of the on-site Coulomb interaction within 
the GGA+$ U $ approach on the electronic structure and especially the 
band gap of the LAO/STO heterostructures using a variety of different 
parameters sets.

\section{Method}

The \textit{ab initio} calculations discussed in the present work were 
based on density functional theory (DFT) \cite{hohenberg1964,kohn1965}. 
Exchange and correlation effects were accounted for by the generalized 
gradient approximation (GGA) as parametrised by Perdew, Burke, and 
Ernzerhof (PBE) \cite{perdew1996}. The Kohn-Sham equations were solved 
with projector-augmented-wave (PAW) potentials and wave functions 
\cite{bloechl1994} as implemented in the Vienna Ab-Initio Simulation 
Package (VASP) \cite{kresse1996,kresse1999}, which is part of the 
MedeA\textsuperscript{\textregistered} software of Materials Design 
\cite{medea}. Specifically, we used a plane-wave cutoff of 400~eV. 
The force tolerance was 0.05~eV/\AA\ and the energy tolerance for the 
self-consistency loop was $ 10^{-5} $~eV. The Brillouin zone of the 
heterostructure was sampled on Monkhorst-Pack grids of 
$ 5 \times 5 \times 1 $ and $ 9 \times 9 \times 1 $ $ {\bf k} $-points 
in order to explore the influence of the density of this mesh; for bulk 
$ {\rm LaAlO_3} $ and $ {\rm SrTiO_3} $ a grid of 
$ 9 \times 9 \times 9 $ $ {\bf k} $-points was used. The GGA+$ U $ 
calculations were performed within the 
simplified approach proposed by Dudarev {\em et al.} \cite{dudarev1998}, 
which takes only the difference of $ U - J $ into account. In the present 
context, we focus on the impact of the $ U - J $--term as applied to the 
La $ 4f $ and the Ti $ 3d $ orbitals on the electronic properties. In 
doing so, we performed a series of spin-degenerate calculations with 
values for $ U $ ranging from zero to 5~eV and 9~eV, respectively, for 
the Ti $ 3d $ and La $ 4f $ states. For the sake of conciseness, we 
will use the symbol $ \bar{U} $ short for $ U - J $ in the text below. 

\begin{figure}[htb]
\centering
\includegraphics[width=1.0\linewidth]{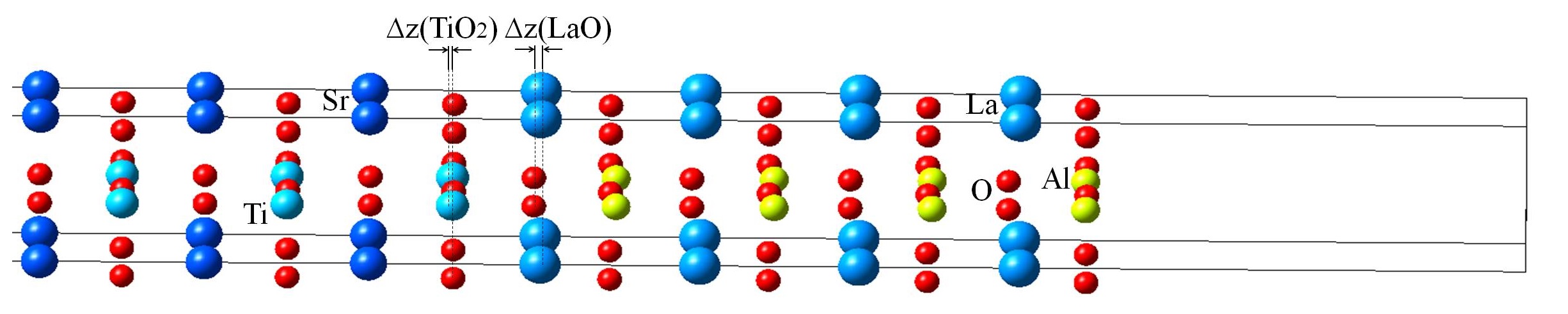}
\caption{Heterostructure consisting of a 4LAO/4.5STO/4LAO slab (see text; 
         only half of the structure is displayed) as arising from a 
         structural optimization using the GGA method. Note the opposite 
         displacements $ \Delta z ({\rm TiO_2}) $ of the $ {\rm Ti}^{4+} $ 
         ions and $ \Delta z ({\rm LaO}) $ of the $ {\rm La}^{3+} $ ions 
         relative to the $ {\rm O}^{2-} $ ions of the respective same 
         layer.} 
\label{fig:structure}
\end{figure}

The heterostructure was modeled by a central region of $ {\rm SrTiO_3} $ 
comprising $ 4 \frac{1}{2} $ unit cells with $ {\rm TiO_2} $ termination 
on both sides and four $ {\rm LaAlO_3} $ overlayers with $ {\rm LaO} $ 
termination towards the central slab and $ {\rm AlO_2} $ surface 
termination also on both sides (see Fig.~4 of Ref.~\cite{piyanzina2016} 
and Fig.~\ref{fig:structure}). 
This slab model guaranteed a non-polar structure without any artificial 
dipoles. In order to avoid interaction of the surfaces and slabs with 
their periodic images a 20~\AA\ vacuum region was added in accordance 
with previous work \cite{cossu2013,piyanzina2016}. The in-plane lattice 
parameter $ a = b = 3.905 $\,\AA\ was fixed to the experimental values of 
bulk $ {\rm SrTiO_3} $ \cite{ohtomo2004} and kept fixed for all subsequent 
calculations reflecting the stability of the substrate. However, the 
atomic positions were fully relaxed. The structural relaxations caused 
particularly displacements of the metal atoms in the layers close to 
the interface relative to the oxygen atoms of the respective layers. 
Interestingly, these displacements point to opposite directions. To be 
specific, a negative displacement $ \Delta z ({\rm TiO_2}) $ of the 
$ {\rm Ti}^{4+} $ ions and a positive displacement $ \Delta z ({\rm LaO}) $ 
of the $ {\rm La}^{3+} $ ions relative to the $ {\rm O}^{2-} $ ions of 
the respective same layer as indicated in Fig.~\ref{fig:structure} were 
obtained. Negative and positive here refer to displacements towards and 
away from the center of the full slab, which, in Fig.~\ref{fig:structure}, 
is located at the left boarder of the plot.

\section{Impact of on-site Coulomb correlations on densities of states}

\begin{figure}[htb]
\centering
\includegraphics[width=0.72\linewidth]{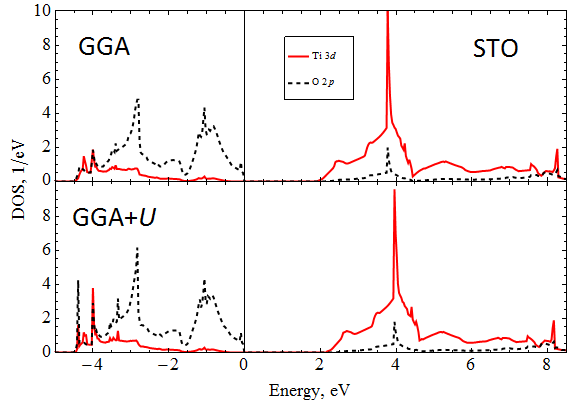} \\
\includegraphics[width=0.72\linewidth]{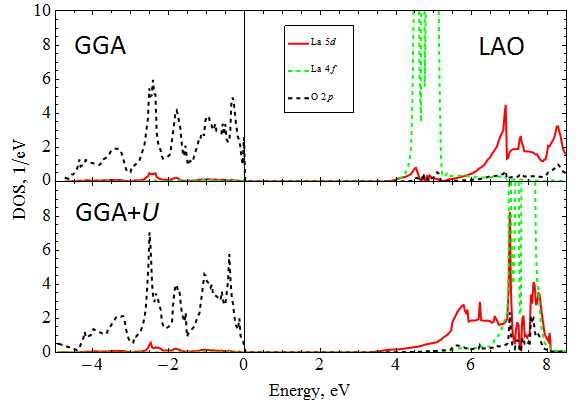}
\caption{Densities of states (DOS) of bulk $ {\rm SrTiO_3} $ (top) 
         and bulk $ {\rm LaAlO_3} $ (bottom) as calculated using the 
         GGA (top) and the GGA+$ U $ method (bottom). $ \bar{U} $ values 
         of 2~eV and 8~eV were used for the Ti $ 3d $ and La $ 4f $ 
         states, respectively.} 
\label{fig:dosbulk}
\end{figure}

\begin{figure}[htb]
\centering
\includegraphics[width=0.72\linewidth]{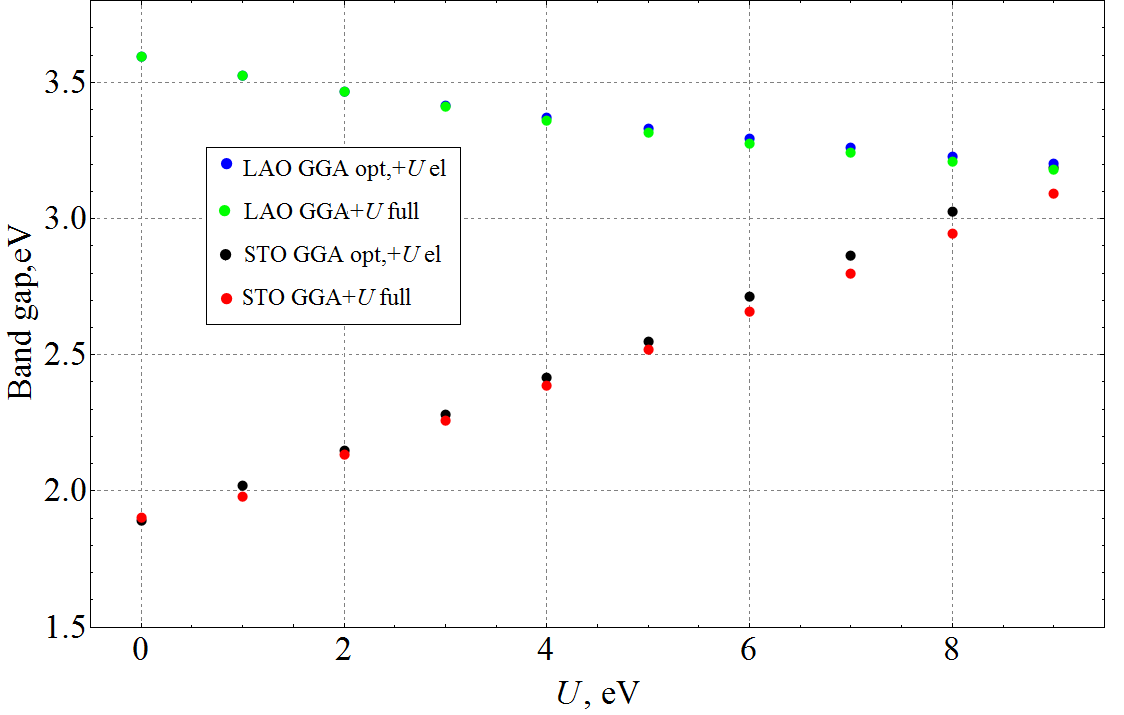}  
\caption{Calculated band gaps of $ {\rm SrTiO_3} $ and $ {\rm LaAlO_3} $ 
         as a function of the parameter $ \bar{U} $. Curves labeled 
         ``GGA opt, +$U$ el'' refer to calculations with the structural 
         relaxation performed at the GGA level and the GGA+$ U $ 
         calculations used only for the calculation of the electronic 
         properties. In contrast, curves labeled ``GGA+$U$ full'' refer to 
         calculations, which used the GGA+$ U $ method in both steps.}
\label{fig:gapbulk}
\end{figure}

In a first step, we investigated the electronic properties of the 
constituent bulk materials, namely, bulk $ {\rm LaAlO_3} $ and 
$ {\rm SrTiO_3} $, and studied their sensitivity to on-site Coulomb 
correlations as captured by the GGA+$ U $ approach. The results are 
displayed in Fig.~\ref{fig:dosbulk}. 
We used $ \bar{U} $ values of 2~eV and 8~eV for the Ti $ 3d $ and 
La $ 4f $ states, respectively. Clearly, on inclusion of on-site 
correlations an upshift of the Ti $ 3d $ states by about 0.25\,eV 
and of the La $ 4f $ states by about 3\,eV relative to the GGA 
results is observed. However, while at the same time the calculated 
band gap of $ {\rm SrTiO_3} $ is increased from 1.9\,eV to 2.15\,eV 
as expected, the band gap of $ {\rm LaAlO_3} $ undergoes a 
considerable counterintuitive decrease from 3.6\,eV to 3.2\,eV. 
According to Fig.~\ref{fig:gapbulk} 
the dependence of the band gap of both bulk systems on $ \bar{U} $ is 
almost linear. Obviously, the band gap depends only slightly on the 
level, at which the GGA+$ U $ method is applied. To be specific, we 
performed two kinds of calculations. In the first set, the GGA+$ U $ 
approach was used for both the structural relaxation and the 
calculation of the band gap. In contrast, in a second set of calculations 
the structural optimization was performed at the GGA level and the GGA+$ U $ 
method used only for the evaluation of the band gap. However, the effect 
of $ \bar{U} $ on the lattice parameter, which is the only structural 
degree of freedom of the perovskite structure, is rather limited. 
To be precise, for $ {\rm SrTiO_3} $ the lattice parameter changes almost 
linearly from about 3.91~\AA\ to 3.96~\AA\ for $ \bar{U}_{\rm Ti} $ ranging 
from 0 to 9\,eV and for $ {\rm LaAlO_3} $ it changes from about 
3.78~\AA\ to 3.80~\AA\ within the same range of $ \bar{U}_{\rm La} $ values. 
These changes are similar to the differences obtained from LDA and GGA 
as well as their deviations from experimentally determined lattice 
parameters, which likewise usually do not have a large impact on the band 
gap. 

Though the decrease of the band gap of bulk $ {\rm LaAlO_3} $ with 
increased values of $ \bar{U} $ comes somewhat unexpected it can be 
understood from closer inspection of the partial densities of states 
shown in Fig.~\ref{fig:dosbulk}. Obviously, the La $ 5d $ states 
experience suppression at energies close to the La $ 4f $ 
states due to strong level repulsion. As a consequence, upshift of the 
La $ 4f $ states due to increase of $ \bar{U} $ causes a suppression 
of the $ 5d $ density of states in the upper part of the $ 5d $ 
bands rather than at their lower edge. At the same time the center of 
gravity of the $ 5d $ bands is shifted downwards by about 1\,eV. 
Since the lower tail of the $ 5d $ partial DOS recombines with the 
main part this downshift is not fully translated into a corresponding 
reduction of the band gap. Instead, the latter is reduced by only 
0.4\,eV as mentioned above. As we will see below, this mechanism is 
likewise effective in the heterostructure, where, however, the 
opposing trends of the band gaps of $ {\rm SrTiO_3} $ and $ {\rm LaAlO_3} $ 
in response to switching on and increasing local correlations as well 
as the presence of both Ti $ 3d $ and La $ 5d $ states will induce a 
more complex behavior. 

\begin{figure}[h!]
\centering $ \bar{U}_{\rm La} = 0 $~eV, $ \bar{U}_{\rm Ti} = 0 $~eV
\vspace{-0.25cm}
\center{\includegraphics[width=0.72\linewidth]{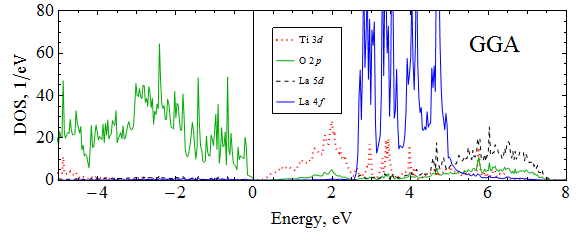}} \\
\centering $ \bar{U}_{\rm La} = 0 $~eV, $ \bar{U}_{\rm Ti} = 2 $~eV
\vspace{-0.25cm}
\center{\includegraphics[width=0.72\linewidth]{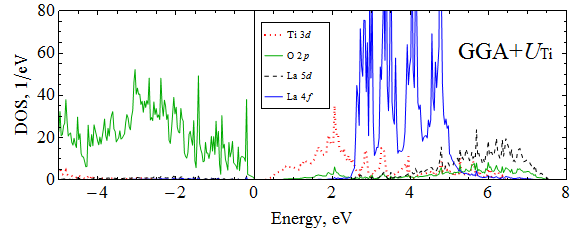} }\\
\centering $ \bar{U}_{\rm La} = 8 $~eV, $ \bar{U}_{\rm Ti} = 0 $~eV
\vspace{-0.25cm}
\center{\includegraphics[width=0.72\linewidth]{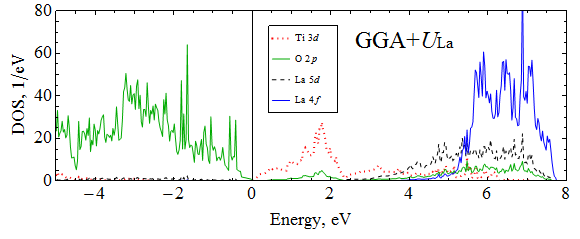} } \\
\centering $ \bar{U}_{\rm La} = 8 $~eV, $ \bar{U}_{\rm Ti} = 2 $~eV
\vspace{-0.25cm}
\center{\includegraphics[width=0.72\linewidth]{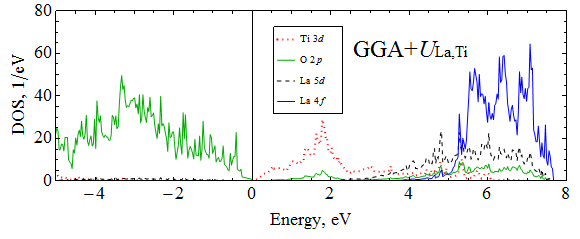}}\\ 
\vspace{-0.3cm}
\caption{Densities of states (DOS) of a 4LAO/4.5STO/4LAO slab 
         as calculated using different values of $ \bar{U} $  
         and a $ {\bf k} $-mesh comprising $ 9 \times 9 \times 1 $ 
         points.}
\label{fig:doshet}
\end{figure}

The total and partial densities of states calculated for a slab consisting 
of the central STO region with four and a half unit cells sandwiched by 
four LAO unit cells on each side are presented in Fig.~\ref{fig:doshet}. 
All calculations employed a $ {\bf k} $-mesh of $ 9 \times 9 \times 1 $ 
points. 
Four different points in the ($ \bar{U}_{\rm La} $, $ \bar{U}_{\rm Ti} $) 
parameter space are considered including GGA calculations and the set 
($ \bar{U}_{\rm La} = 8 $~eV, $ \bar{U}_{\rm Ti} = 2 $~eV). For all sets, 
we observe dominant O $ 2p $ valence bands, which are (almost completely) 
separated from the metal $ d $ and $ f $ states. Especially the 
latter experience considerable shifts and deformations as soon as local 
correlations via the $ \bar{U} $ parameters are taken into account. To  
be specific, within the GGA semiconducting behavior with a small band 
gap of 0.13~eV is observed, which turns into metallic conductivity on the 
addition of a fifth LAO layer. This is in agreement with previous 
studies~\cite{nazir2014}. Worth mentioning is the overall width of the 
La $ 4f $ bands, which results from a distribution of single peaks of the 
various La atoms in the different $ {\rm LaO} $ layers. 
As expected, on adding a $ \bar{U} $ of 2~eV at the Ti sites, the band gap 
increases to a value of 0.25~eV while the overall shape of the partial 
densities of states is not affected. This changes drastically as soon as 
local correlations for the La $ 4f $ states are taken into account. In 
particular, the $ 4f $ states shift by about 3\,eV just as observed for 
bulk $ {\rm LaAlO_3} $. In addition, $f$-wave bands originating from
atoms in different planes are concentrated in a narrower energy range. 
Most importantly, however, a downshift of the 
Ti $ 3d $ bands is observed, which in turn leads to metallic band 
overlap for both $ \bar{U}_{\rm Ti} = 0 $~eV and $ \bar{U}_{\rm Ti} = 2 $~eV. 
While the reduction of the band gap is similar to the behavior observed 
for $ {\rm LaAlO_3} $, it is now primarily due to the Ti $ 3d $ states, 
which form the lower edge of the conduction band. Nevertheless, we still 
find considerable downshift of the center of gravity of the La $ 5d $ 
states with increasing $ \bar{U}_{\rm La} $. This is mainly due to the 
suppression of these states in the energy window between about 3 and 
4.5~eV observed in the calculations with zero $ \bar{U}_{\rm La} $. 
In contrast, the upper part of the La $ 5d $ partial DOS between 5 and 
7.5~eV is rather unaffected by the shift of the La $ 4f $ states. 
Restoration of the La $ 5d $ states in the range from 3 and 4.5~eV due 
to the upshift of the La $ 4f $ levels induced by the increased 
$ \bar{U}_{\rm La} $ leads to a concomitant downshift of the 
Ti $ 3d $ bands and, hence, to closing of the band gap. We will discuss 
possible origins of this behavior below. 

The partial densities of states also reveal an overall similarity 
within the ($ \bar{U}_{\rm La} $, $ \bar{U}_{\rm Ti} $) parameter 
space. As even closer inspection has shown, in all cases crystal 
field splitting of the Ti $ 3d $ states into weakly bonding $ t_{2g} $ 
and well separated bonding and antibonding $ e_g $ states is found. 
As a consequence, 
the $ t_{2g} $ manifold is forming the conduction band minimum at and 
close to the interface. Furthermore, we observe additional splitting 
of these states into low lying interplanar $ d_{xy} $ states as well as 
$ d_{xz} $ and $ d_{yz} $ states at slightly elevated energies in 
agreement with previous work \cite{zhong2013,zabaleta2016}. In contrast, 
the latter two states become more pronounced at the conduction band 
minimum in $ {\rm TiO_2} $ layers away from the interface. A 
layer-resolved analysis of the evolution of the $ t_{2g} $ states 
within the central STO region of up $ 20 \frac{1}{2} $ layer thickness 
has been provided in Ref.~\cite{zabaleta2016}.

\section{Layer-dependence of formation of the metallic interface}

\begin{figure}[htb]
\centering
\includegraphics[width=0.96\linewidth]{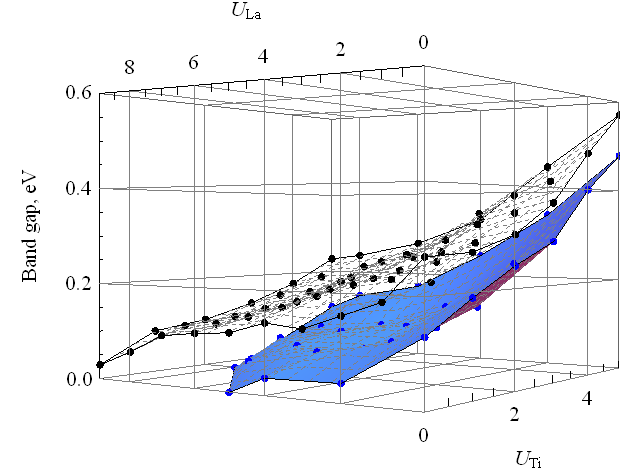}  
\caption{Calculated band gap as a function of $ \bar{U}_{\rm La} $ 
         and $ \bar{U}_{\rm Ti} $. The black and blue points refer 
         to calculations using {\bf k}-point grids of 
         $ 5 \times 5 \times 1 $ and $ 9 \times 9 \times 1 $ points, 
         respectively. The dashed black and solid blue surfaces result 
         from interpolation between the calculated band gaps and serve 
         as guides to the eyes only.} 
\label{fig:gaphet}
\end{figure}

Motivated by the previous results and due to the lack of detailed 
studies in the literature we complemented the previous results by 
a systematic investigation of the dependence of the electronic 
properties on the values of $ \bar{U}_{\rm La} $ and 
$ \bar{U}_{\rm Ti} $. The results of this screening are summarized 
in Fig.~\ref{fig:gaphet}, 
which displays the calculated band gaps for a variety of combinations 
of the two $ \bar{U} $ values. The results are arranged on two sheets 
corresponding to the above-mentioned two 
different $ {\bf k} $-point grids with $ 5 \times 5 \times 1 $ and 
$ 9 \times 9 \times 1 $ points, respectively. Obviously, the band gap 
changes by about 0.1~eV on going from the coarse grid to the fine grid. 
However, this change is almost independent of the particular values of 
$ \bar{U}_{\rm La} $ and $ \bar{U}_{\rm Ti} $. Both sheets thus reflect 
the findings already discussed above, namely, the strong increase of 
the band gap with increasing $ \bar{U}_{\rm Ti} $ and its decrease with 
increasing $ \bar{U}_{\rm La} $. 

\begin{figure}[htb]
\centering
\includegraphics[width=0.96\linewidth]{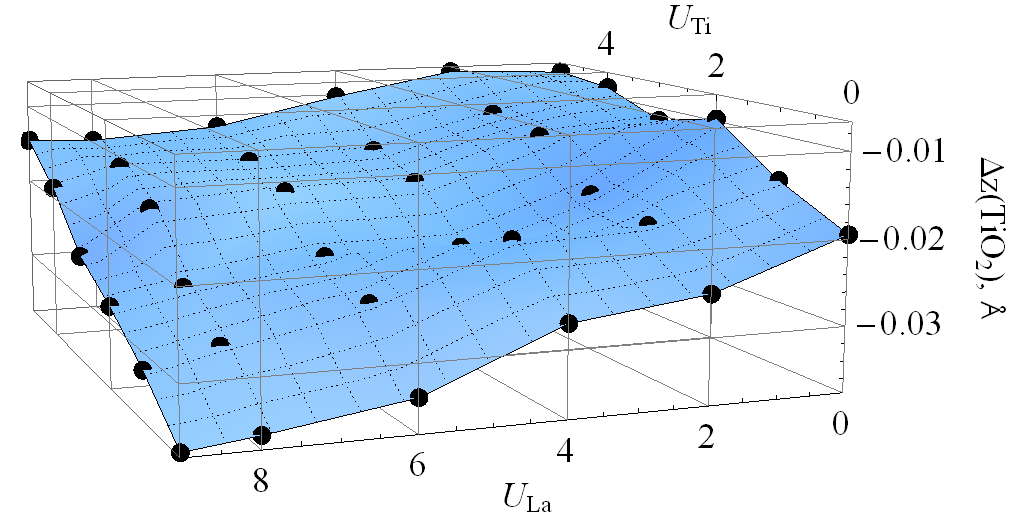}  
\caption{Displacement $ \Delta z ({\rm TiO_2}) $ of Ti atoms out of the 
         $ {\rm TiO_2} $-plane next to the interface as a function of 
         $ \bar{U}_{\rm La} $ and $ \bar{U}_{\rm Ti} $. The solid blue 
         surface results from interpolation between the calculated 
         points and serves as a guide to the eyes only.} 
\label{fig:displ-Ti}
\end{figure}

As already discussed in connection to Fig.~\ref{fig:doshet}, interpretation 
of the opposing response of the band gap to the $ \bar{U} $ parameters 
in terms of the partial densities of states and level repulsion between 
the $ d $ and $ f $ states is not as clear as for bulk $ {\rm LaAlO_3} $. 
We have thus inspected other calculated quantities of the heterostructure 
in order to identify a possible origin of the band-gap behavior, especially
in dependence on $ \bar{U}_{\rm La} $. Indeed, 
a possible explanation is offered by the displacement of the Ti atoms 
out of the oxygen plane at the interface as a function of 
$ \bar{U}_{\rm La} $ and $ \bar{U}_{\rm Ti} $. According to 
Fig.~\ref{fig:displ-Ti} 
this displacement shows the opposite dependence on the $ \bar{U} $ values 
as the band gap, namely, a (negative) decrease with increasing 
$ \bar{U}_{\rm Ti} $ and a (negative) increase with increasing 
$ \bar{U}_{\rm La} $. The relation to the band gap can thus be understood 
in terms of a simple chemical picture based on the bonding-antibonding 
splitting of the Ti $ 3d $ and O $ 2p $ orbitals. A short distance 
between these two types of atoms as signaled by a small (negative) 
displacement leads to a large band gap and vice versa as is indeed 
observed in Fig.~\ref{fig:gaphet}. 

\begin{figure}[t]
\centering $ \bar{U}_{\rm La} = 0 $~eV, $ \bar{U}_{\rm Ti} = 2 $~eV
\vspace{-0.25cm}
\center{\includegraphics[width=0.72\linewidth]{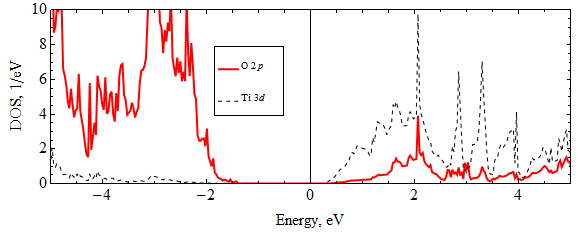}} \\
\centering $ \bar{U}_{\rm La} = 8 $~eV, $ \bar{U}_{\rm Ti} = 2 $~eV
\vspace{-0.25cm}
\center{\includegraphics[width=0.72\linewidth]{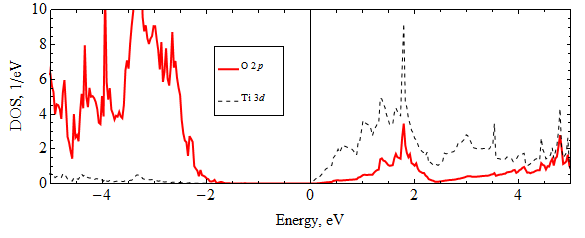}}\\ 
\vspace{-0.3cm}
\caption{Partial densities of states (DOS) of a 4LAO/4.5STO/4LAO slab 
         as calculated using the sets  
         ($ \bar{U}_{\rm La} = 0 $~eV, $ \bar{U}_{\rm Ti} = 2 $~eV) and 
         ($ \bar{U}_{\rm La} = 8 $~eV, $ \bar{U}_{\rm Ti} = 2 $~eV). 
         The partial densities of states comprise contributions from 
         atoms in the interface layer only.}
\label{fig:doshet-if}
\end{figure}

In order to check if this scenario can indeed explain the change of 
the band gap we calculated the partial densities of states of the 
atoms of the $ {\rm TiO_2} $ layer at the interface, i.e.\ of those 
atoms, which are covered by Fig.~\ref{fig:displ-Ti}, for the sets 
($ \bar{U}_{\rm La} = 0 $~eV, $ \bar{U}_{\rm Ti} = 2 $~eV) and 
($ \bar{U}_{\rm La} = 8 $~eV, $ \bar{U}_{\rm Ti} = 2 $~eV). If the 
above scenario were valid, an increased value of $ \bar{U}_{\rm La} $ 
should via an increased absolute value of the displacement 
$ \Delta z ({\rm TiO_2}) $ of Ti atoms out of the $ {\rm TiO_2} $ 
plane induce a decreased energetical separation of the O $ 2p $ and 
Ti $ 3d $ bands. However, the calculated partial densities of states 
as displayed in Fig.~\ref{fig:doshet-if} 
do not give any indication for such a scenario. Indeed, on going from 
$ \bar{U}_{\rm La} = 0 $~eV to $ \bar{U}_{\rm La} = 8 $~eV the O $ 2p $ 
bands are shifted downwards by the same amount as the Ti $ 3d $ levels. 
We thus have to rule out the displacement of the Ti atoms out of the 
$ {\rm TiO_2} $ plane at the interface as the major source of the 
decreasing band gap with increasing $ \bar{U}_{\rm La} $. 

\begin{figure}[htb]
\centering
\includegraphics[width=0.96\linewidth]{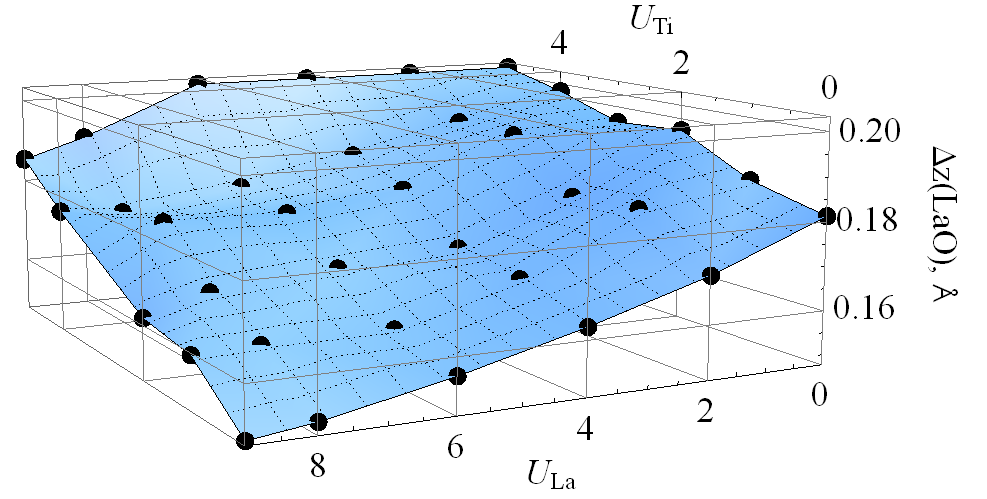}  
\caption{Displacement $ \Delta z ({\rm LaO}) $ of La atoms out of the 
         LaO-plane next to the interface as a function of 
         $ \bar{U}_{\rm La} $ and $ \bar{U}_{\rm Ti} $. The solid blue 
         surface results from interpolation between the calculated 
         points and serves as a guide to the eyes only.} 
\label{fig:displ-La}
\end{figure}

\begin{figure}[tb]
\centering
\includegraphics[width=0.72\linewidth]{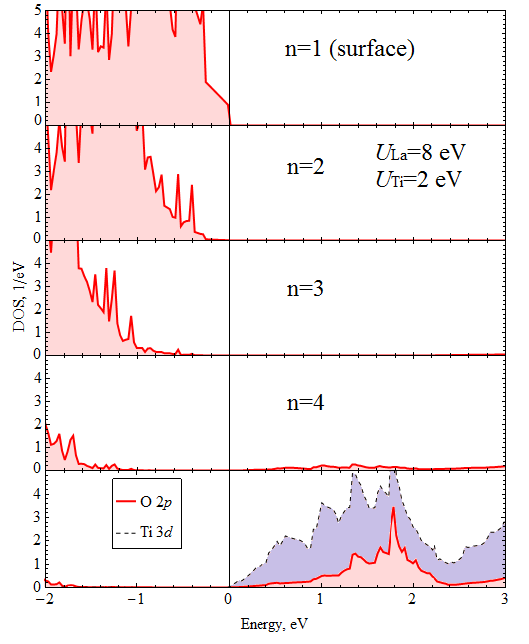}  
\caption{Partial densities of states (DOS) of a 4LAO/4.5STO/4LAO slab 
         as calculated using the set 
         ($ \bar{U}_{\rm La} = 8 $~eV, $ \bar{U}_{\rm Ti} = 2 $~eV). 
         The partial DOS comprise contributions from atoms inside 
         the indicated layers only (see also 
         Refs.~\cite{pentcheva2009,schwingenschloegl2009,pavlenko2011}).}
\label{fig:doshet-layer}
\end{figure}

Next, we turn to a different scenario, which is based on the 
displacement $ \Delta z ({\rm LaO}) $ of La atoms out of the LaO 
plane at the interface as a function of the $ \bar{U} $ parameters 
as displayed in Fig.~\ref{fig:displ-La}.  
For all values of the $ \bar{U} $ parameters included in the figure 
we find a positive displacement, which signifies that the La atoms 
are shifted away from the LaO interface layer, as shown in Fig.~1b of 
Ref.~\cite{zabaleta2016}. Most prominently, the shift is reduced for
increasing $ \bar{U}_{\rm La} $, which according to 
Fig.~\ref{fig:displ-Ti} is accompanied by a negative displacement of 
the Ti atoms in the interface layer as is expected when electrostatics 
is accounted for. This scenario was already addressed by Zabaleta 
{\it et al.}\ in the context of the application of hydrostatic pressure, 
which likewise tends to reduce the buckling in the LaO layer 
\cite{zabaleta2016}. According to Zabaleta {\it et al.}, the concomitant 
suppression of the dipole moment of this layer would reduce the screening 
of the polar discontinuity, which fact enhances the charge density at 
the interface \cite{zabaleta2016}. In the present context we thus suggest 
that increasing $ \bar{U}_{\rm La} $, via the reduced dipole moment in 
the LaO layer and the resulting reduced screening of the polar 
discontinuity, eventually entails a corresponding increase of the 
electrostatic potential across the LAO slab. The latter induces  
an upshift of the surface O $ 2p $ states relative to the Ti $ 3d $ states 
at the interface that, according to the layer-resolved partial 
densities of states displayed in Fig.~\ref{fig:doshet-layer} 
(see also Refs.~\cite{pentcheva2009,schwingenschloegl2009,pavlenko2011}), 
closes the band gap between these states. 
We thus arrive at the important conclusion that the effect especially 
of $ \bar{U}_{\rm La} $ on the electronic properties is only indirect 
since this parameter mainly controls the buckling of the $ {\rm LaO} $ 
planes and of the TiO$_2$ interface plane and affects the electronic bands
only via this structural characteristic.

\section{Conclusions}

The present study has led to a variety of important results. While 
for bulk $ {\rm SrTiO_3} $ increasing influence of local electronic 
correlations as captured by the on-site Coulomb repulsion parameter 
$ \bar{U}_{\rm Ti} $ applied to the Ti $ 3d $ states leads to the 
expected increase of the band gap, a counterintuitive decrease of 
the band gap with increasing value of $ \bar{U}_{\rm La} $ for the 
La $ 4f $ states of bulk LaAlO$_3$ is observed. From analysis of the partial densities 
of states we were able to explain this behavior by recovery of the 
lower conduction band of La $ 5d $ character from deformation due to 
$ 5d $-$ 4f $ level repulsion as the La $ 4f $ states shift to higher 
energies and thereby to identify the purely electronic origin of this 
behavior. 
While for the heterostructure, analysis of the partial densities of 
states likewise reveals increase and decrease or even closure of the 
band gap with increasing $ \bar{U}_{\rm La} $ and $ \bar{U}_{\rm Ti} $, 
respectively, as well as the $ 5d $-$ 4f $ level repulsion, a simple 
interpretation is spoiled by the presence of the Ti $ 3d $ states at 
the conduction band edge.

Systematic screening of the 
($ \bar{U}_{\rm La} $, $ \bar{U}_{\rm Ti} $) parameter space allowed 
to identify correlations between different structural and electronic 
properties. In particular, decrease/increase of the band gap goes 
along with (negative) increase/decrease of the displacement of the 
Ti atoms out of the $ {\rm TiO_2} $ interface plane and (positive) 
increase/decrease of the La atoms out of the $ {\rm LaO} $ interface 
plane. However, inspection of the partial densities of states gave 
no indications for a change in bonding-antibonding splitting between 
the O $ 2p $ and Ti $ 3d $ bands, which would explain the change in 
band gap in terms of the structural distortion with the $ {\rm TiO_2} $ 
interface layer. Following the pressure study of 
Zabaleta {\em et al.}~\cite{zabaleta2016} 
we thus arrive at a non-local scenario relating the band-gap trends 
to the buckling of the LaO layers close to the interface, which in 
turn has a strong impact on the dipole moment of this layer. Reduced 
buckling would thus translate into reduced screening of the polar 
discontinuity and a stronger electric field in the LAO slab, which causes 
upshift of the O $ 2p $ valence states at the surface relative to the  
Ti $ 3d $ conduction bands at the interface and thus decrease of the 
band gap. In conclusion, our calculations reveal that the impact of 
local electronic correlations on the electronic properties is 
primarily due to strong electron-lattice coupling, which controls 
the buckling of the LaO layers close to the interface.

\section*{Acknowledgments}

This study was supported by the Supercomputing Center of Lomonosov Moscow 
State University. The authors from KFU acknowledge partial support by the 
Program of Competitive Growth of Kazan Federal University. I.\,Piyanzina 
received financial support from the German Academic Exchange Service (DAAD) 
as well as from the German Science Foundation (DFG) through the transregional 
collaborative research center TRR\,80.

\end{document}